\documentclass[runningheads,a4paper]{llncs}

\usepackage{amssymb}
\usepackage{graphicx}
\graphicspath{ {./figures/} }
\usepackage{url}

\usepackage[llncs,crop]{llncsconf}
\conference{International Conference on \LaTeX-Hacks}
\llncs{This is a post-peer-review, pre-copyedit version of an article published in Lecture Notes in Computer Science (LNCS, volume 8696).}{1}
\llncsdoi{10.1007/978-3-319-10557-4_43}

\urldef{\mailsa}\path|{marc.zeller, kai.hoefig, martin.rothfelder}@siemens.com|

\begin{document}

\mainmatter  

\title{Towards a Cross-domain Software Safety Assurance Process for Embedded Systems}

\titlerunning{Cross-domain Software Safety Assurance Process for Embedded Systems}

%
%
\author{Marc Zeller, Kai H{\"o}fig, Martin Rothfelder}

\institute{Siemens AG, Corporate Technology\\
Otto-Hahn-Ring 6, 81379 Munich, Germany\\
\mailsa\\
}

%
%

\maketitle

\begin{abstract}
In this work, we outline a cross-domain assurance process for safety-relevant software in embedded systems. This process aims to be applied in various different application domains and in conjunction with any development methodology. With this approach we plan to reduce the growing effort for safety assessment in embedded systems by reusing safety analysis techniques and tools for the product development in different domains.
\end{abstract}

\section{Introduction}
\label{sec:introduction}
%
The importance of safety-relevant software systems in many application domains of embedded systems, such as aerospace, railway, health care, automotive and industrial automation, is continuously growing.
Thus, along with the growing system complexity, also the need for safety assessment as well as its effort is increasing drastically in order to guarantee the high quality demands in these application domains. However, this trend is contrary to industry's aim to reduce development costs and time-to-market of new products.

The goal of safety assessment is to identify all failures that cause hazardous situations and to demonstrate that their probabilities are sufficiently low. In the application domains of safety-relevant software the safety assurance process is defined by the means of safety standards. The requirements of these standards must be met in order to enable argumentation that the system is safe.
To reduce development costs and the time-to-market, one possible approach is to develop a safety assurance process which is applicable to multiple applications domains of embedded systems (e.g. like the IEC 61508 standard \cite{iec61508}).
In this paper, we present an approach towards a safety assurance process for software which is applicable across different application domains of embedded systems. This process aims to be applicable with various development methodologies used in different domains and tries to use common safety analysis techniques as far as possible.
Hence, it builds the foundation for the future development of methods and tools for safety assurance which can be applied across domains of safety-relevant software systems.
Thus, safety analysis techniques and tools as well as artifacts produced during the safety assurance process may be reused for the safety assessment of different kinds of products. Especially, in areas where embedded systems are highly related to software product-lines or heterogeneous systems-of-systems, a cross-domain safety analysis process will reduce the effort needed to fulfill the requirements of the respective safety standard significantly. 

This paper is organized as follows: In Sec.~\ref{sec:relatedWork} we present relevant related work. Then, we outline our approach for a cross-domain safety assurance process. The benefits of this process are discussed in Sec.~\ref{sec:discussion}. This paper is concluded and an outlook to future work is given in the last section.

\section{Related Work}
\label{sec:relatedWork}
Today, numerous standards related to functional safety of software are existing (cf.~\cite{baufreton2010multi}).
These standards provide the rules and guidelines as basis for the safety assurance process of safety-relevant systems in specific domains. Since each domain-specific safety standard defines a specific vocabulary and covers the complete safety life-cycle, each domain has evolved its individual safety assurance process.
Since we focus on the safety assurance process of software, we will further consider only safety standards related to software. These standards are: the DO-178C in aeronautics, the ISO 26262 in automotive, IEC 60880 \& 62138 in the area of nuclear power plants, the EN 50128 in railway, the IEC 62304 in health care, and the ECSS-Q-ST-80C in space. Moreover, IEC 61508 will be considered which covers the industrial automation domain. However, only parts of the safety process defined in the IEC 61508 standard are related to software, since the scope of this standard is much broader.

In order to enable cross-domain harmonization of the safety assurance process and the sharing of common techniques and tools, first attempts to identify similarities and dissimilarities have already been performed in \cite{baufreton2010multi,blanquart2012criticality,ledinot2012cross,machrouh2012cross,papadopoulos1999potential}.
As a result of these previous analyses the following similarities between the examined standards have be identified:
\begin{itemize}
	\item Common notion of safety and certification
  \item Linear progressing safety process with dedicated phases
  \item Combined hazard assessment and risk analysis to derive safety requirements
  \item Criticality levels as means to allocation safety (integrity) requirements to system elements
  \item Verification activities are driven by the safety requirements
  \item Safety case provides evidence that safety requirements are fulfilled which is needed for certification
\end{itemize}
Moreover, the following divergences have be identified:
\begin{itemize}
  \item Varying definition of criticality levels
  \item Different approaches for the allocation of safety requirements
  \item Specific verification \& validation processes
\end{itemize}

Based on a number of identified similarities of the safety standards in the transportation sector, \cite{papadopoulos1999potential} already outlines a generic safety assessment process integrated into a concrete system development process.
However, only safety standards from the transportation domain are analyzed and recent developments in safety regulations (e.g. the ISO 26262) are not considered. In our work, we aim at finding a cross-domain safety assurance process applicable to any domain-specific software development process.

\section{Cross-domain Safety Assurance Process}
\label{sec:genericProcess}
\begin{figure}
  \centering
  \includegraphics[width=0.55\textwidth]{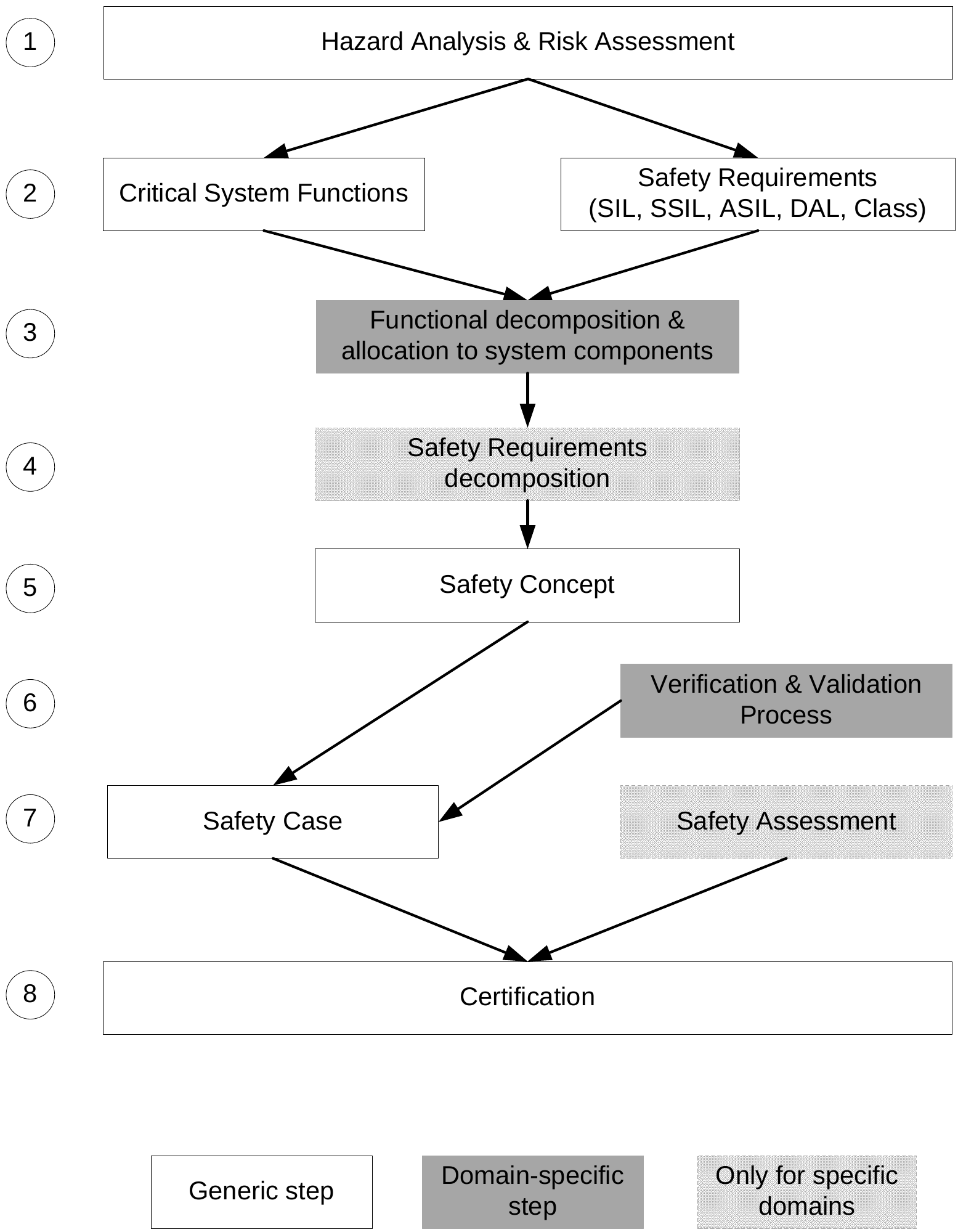}
  \caption{Cross-domain software safety assurance process}
  \label{fig:process}
\end{figure}

According to the similarities and divergences of the analyzed safety standards related to software, we outline a cross-domain safety assurance process (see Fig.~\ref{fig:process}). This process consists of generic and domain-specific steps which must be executed in each of the considered domains as well as steps which are only necessary in specific domains.

The software safety assurance process starts with the \textit{Hazards Analysis \& Risk Assessment (HARA)}. This step aims to determine safety-relevant systems functions, the safety requirements of these functions (maximum tolerable failure probabilities) as well as the potential demands for additional safety functions. The different safety standards generally agree on common HARA techniques \cite{machrouh2012cross}.

As a result of the HARA, safety requirements are derived.
According to \cite{baufreton2010multi}, all safety standards introduce criticality categories, so-called \textit{Criticality Levels} (e.g. the Safety Integrity Levels (SILs) in IEC 61508) to quantify the safety requirements.
In all domains, the criticality levels characterize the consequences of failures (severity) combined with a notion of their occurrence probability.
However, in each domain the acceptability frontiers of risk differs due to divergences in the definition of severity and risk occurrence \cite{blanquart2012criticality}.

The next phase in the assurance process is the allocation of criticality levels to system elements as well as the functional decomposition.
This step is domain-specific due to the differences between safety standards. In aeronautics, nuclear, railway and space the fist categorized element is a (top level) function. Then the criticality levels are derived according to functional decomposition and allocated to the system elements implementing the top-level functions. In the automotive domain, however the criticality levels are allocated first to safety goals and derived to safety requirements and system elements.

Furthermore, a decomposition of the criticality levels is possible in railway and space domain (according to generic rules) as well as in the aeronautics and automotive domain (according to specific rules) \cite{blanquart2012criticality}. But this phase of the safety assurance process is only possible in these domains.

As a next step, the \textit{Safety Concept} is derived. It is defined as the specification of the safety requirements, their allocation to system elements and their interaction necessary to achieve safety goals \cite{iso26262}. The construction of the safety concept is a generic step which is compliant to all considered safety standards. However, the necessary content of the safety concept may differ from domain to domain.

Within the next phase, the developed software is verified and validated. The verification and validation techniques (such as source code verification, unit testing, integration testing, etc.) and the process itself are defined or recommended by the domain-specific safety standards.

Finally, the \textit{Safety Case} is compiled to argue that the system is safe. The safety case is derived from the safety concept and extended by results of the verification \& validation process to prove that the safety requirements have been fulfilled.
A safety case is a concept applicable across all domains. However, in aerospace and railway a so-called \textit{Safety Assessment} is required additionally to show the conformity to the standard \cite{ledinot2012cross}.
The evidence for system safety provided by the safety case forms the basis for safety certification.

\section{Discussion}
\label{sec:discussion}
The cross-domain safety assurance process outlined in Sec.~\ref{sec:genericProcess} is applicable to any development process or methodology,
since none of the process steps directly include a reference to the system development.
Hence, our approach may be used along with state-of-the-art approaches in system/software engineering such as component-based and model-based development as well as software-product-lines.

Our approach solely consists of 8 different process steps which are all not new in the area of safety engineering. Thus, established safety analysis techniques (such as \textit{Hazard and Operability Studies (HAZOP)}, \textit{Failure Mode and Effects Analysis (FMEA)}, \textit{Fault Tree Analysis (FTA)}, etc.) can be applied and no new safety analysis methodologies need to be developed.

More than the half of the phases in our process are generic. Therefore, the techniques and tools used in these steps can be applied in various application domains. The domain-specific steps only differ in methods for the allocation of criticality levels to the system elements and the requirements for software verification \& validation. However, also common techniques and tools can be applied in these phases but have to be adapted to the domain-specific requirements respectively. There are only two process phases which are solely relevant to particular safety standards. For these steps domain-specific techniques and tools must be provided separately.

\section{Conclusions and Outlook}
\label{sec:conclusion}
%
The cross-domain assurance process for safety-relevant software in embedded systems, outlined in this paper, aims to be applied in various different application domains. Thus, supporting the cost-efficient system development as well as the reuse of techniques and tools for the safety analysis.
However, not all of the process steps can be realized in a generic and domain-independent way.
But our approach is independent from concrete development methodologies and can be applied along with component-based and model-based design. Moreover, common safety analysis techniques can by applied in most process steps.
  
Future work will include a more refined description of each phases of the cross-domain safety assurance process including applicable techniques and tools. Moreover, we will evaluate our approach in different application domains.

\bibliography{references}
\bibliographystyle{splncs03}

\end{document}